# Investigating $^{16}$O via the $^{15}$N(*p*,α)$^{12}$C reaction

J.A. Swartz[1,a], H.O.U. Fynbo[1,b], K.L. Andersen[1,c], M. Munch[1] and O.S. Kirsebom[1]

[1]*Department of Physics and Astronomy, Aarhus University, Ny Munkegade 120, Aarhus C, Denmark*

[a)]jasw@dtu.dk
[b)]fynbo@phys.au.dk
[c)]kenneth@phys.au.dk

**Abstract.** The $^{16}$O nucleus was investigated through the $^{15}$N(*p*,α)$^{12}$C reaction at excitation energies from $E_x$ = 12 231 to 15 700 keV using proton beams from a 5 MeV Van de Graaff accelerator at beam energies of $E_p$ = 331 to 3800 keV. Alpha decay from resonant states in $^{16}$O was strongly observed for ten known excited states in this region. The candidate 4-alpha cluster state at $E_x$ = 15.1 MeV was investigated particularly intensely in order to understand its particle decay channels.

## PHYSICS BACKGROUND

The target nucleus of this investigation, $^{15}$N, is of interest to nuclear astrophysics as it acts as a branching point nucleus in the CNO cycle. Through the $^{15}$N(*p*,α)$^{12}$C reaction it may create $^{12}$C and thus serve as a trigger to the CNO 1 cycle, while through the $^{15}$N(*p*,γ)$^{16}$O reaction it may start the CNO 2 cycle [1]. In either case, the reaction proceeds through resonant states of $^{16}$O and requires the participation of a proton. The astrophysical S-factor of the $^{15}$N(*p*,α)$^{12}$C reaction is thought to be influenced by two known resonant 1$^-$ states at 12.44 and 13.09 MeV in $^{16}$O [2,3]. This reaction was recently investigated experimentally and described with a R-matrix formalism by the Notre Dame group up to an $^{16}$O excitation energy of $E_x$ = 13.5 MeV [3,4]. The present measurement may enable for a description of this reaction channel beyond the 4-alpha breakup threshold (at $E_x$ = 14.4 MeV in $^{16}$O), up to $E_x$ = 15.7 MeV, which may also improve the R-matrix description at the astrophysically relevant lower excitation energies. This region also contains candidates for the 4-alpha cluster state such as the $0_4^+$ state at $E_x$ = 13.6(1) MeV [5] and the $0_6^+$ state at $E_x$ = 15.1 MeV [6,7]. The latter state was recently investigated through $^{16}$O(α,α')$^{16}$O inelastic scattering at iThemba LABS, and it was found that another state, exhibiting a very strong alpha-1 decay strength, may exist about 50 keV below it [8]. It is possible that this state has obscured previous measurements of the $0_6^+$ state. We attempt here to resolve this question and to compare the $^{15}$N(*p*,α)$^{12}$C and $^{15}$N(*p*,γ)$^{16}$O strengths in all states observed from $E_x$ = 12 – 16 MeV in $^{16}$O.

## EXPERIMENTAL METHOD

The Aarhus University 5 MeV Van de Graaff accelerator was used to provide a proton beam at $E_p$ = 700 - 3800 keV to a target of C$^{15}$N$_x$ which was surrounded by a closely-packed array of double-sided Silicon strip detectors (DSSDs), shown in Fig 1 (b) and (c). Two annular S3 detectors were placed up and downstream from the target, while two square wedge-shaped detectors were placed at right angles to the beam on both sides of and at about 4 cm from the target. The C$^{15}$N$_x$ target was produced by radio frequency magnetron sputtering (RF MS), with a thickness of 40 μg/cm$^2$ and a composition of 40% of $^{15}$N, thus yielding an effective $^{15}$N thickness of about 16 μg/cm$^2$. This detection setup yields a total solid angle coverage of about 40%. Through measurement over long periods of beamtime, excited levels in nuclei can be investigated extensively with this setup.

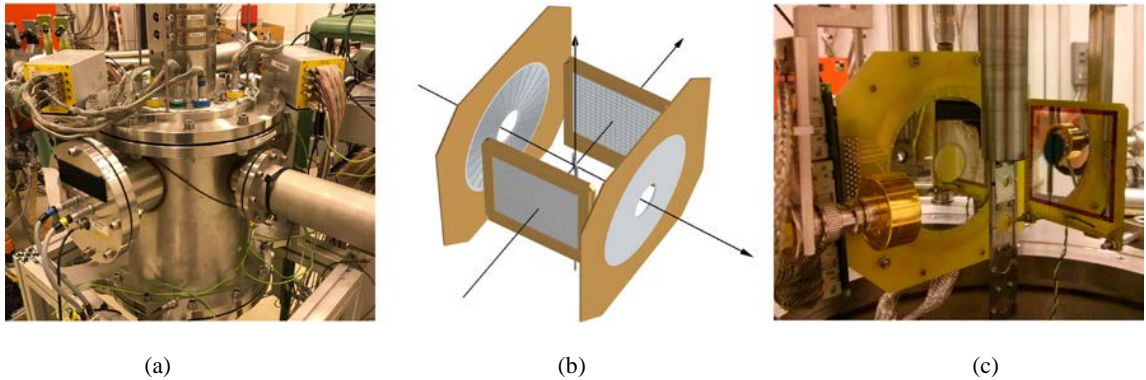

(a)          (b)          (c)

**FIGURE 1.** The target chamber is shown from outside in (a), and the DSSD detection array is presented by the diagram in (b) and partly by the photo in (c), which includes the top four out of five slots of the target ladder [9]

## ANALYSIS AND RESULTS

Kinematic reconstruction of events where $^{12}$C and alpha particles were observed in coincidence in the DSSD detectors was performed to obtain excitation energy information for the $^{16}$O compound nucleus for events related to the $^{15}$N$(p,\alpha)^{12}$C reaction channel. Thus, $^{16}$O excitation energy spectra such as Fig. 2 could be produced. Fig. 2 also contains peaks between the two alpha channels at around 12 MeV. It was initially speculated that these events may be related to the $^{15}$N$(p,\gamma)^{16}$O reaction channel, where the alpha particle from the subsequent $^{12}$C + α breakup will have less energy due to the emission of the gamma ray. This would give an indirect signal of gamma decay to $^{16}$O, possibly to its 12.049 MeV $0^+$ and 11.096 MeV $4^+$ states. Further investigation, however, revealed these loci to be associated rather with the $^{19}$F$(p,\alpha)^{16}$O reaction channels to the 6.049 MeV $0^+$, 6.917 MeV $2^+$ and 7.116 MeV $1^-$ states in $^{16}$O. The reason for the presence of fluorine on the target is presently unknown. By comparing the strengths of these fluorine-related peaks to those of the $\alpha_0$ and $\alpha_1$ peaks, one may estimate the relative strengths of these hypothetical gamma decays from the 15.1 MeV $0_6^+$ state to the 12.049 MeV and 11.096 MeV states in $^{16}$O, and from literature values deduce the maximum values for the gamma reduced widths, as was done in Table 1.

The yield of $^{12}$C + α events was extracted and normalized to integrated beam current for every proton energy which was investigated by the energy scan. This enabled for the production of the normalized yield to excitation energy spectra for $^{15}$N$(p,\alpha_0)^{12}$C and $^{15}$N$(p,\alpha_1)^{12}$C events which are shown in Fig. 3. At Ex = 3 – 3.4 MeV, a number of data points were shifted 30 keV lower in excitation energy in order to fit with the rest of the data. This was done to compensate for a 30 keV upward shift which is seen in all data acquired from 20 to 28 March 2018. The reason for this shift cannot be explained at present. For more details on the data points where this shift was applied, refer to Ref. 9.

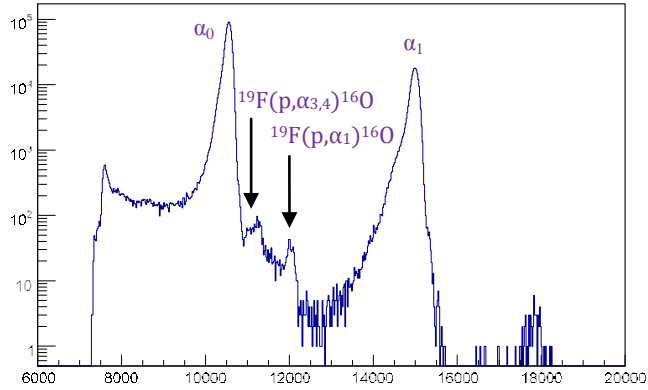

**FIGURE 2.** Spectrum of $^{16}$O excitation energy from 32 hours of data taking with a proton beam incident upon the C$^{15}$N$_x$ target. The $^{15}$N$(p,\alpha_0)^{12}$C and $^{15}$N$(p,\alpha_1)^{12}$C events are indicated by $\alpha_0$ and $\alpha_1$.

**Table 1.** Maximum gamma reduced widths inferred from comparing the strengths of the $^{19}$F$(p,\alpha)^{16}$O channels, which coincided with the kinematic regions of the gamma channels, to literature values of the $^{15}$N$(p,\alpha)^{12}$C channel strengths.

| $^{16}$O excitation level | Ratio γ/a$_0$ strength | Ratio γ/a$_1$ strength | Inferred maximum γ reduced width |
| --- | --- | --- | --- |
| 12.049 MeV 0$^+$ | 0.00256 | 0.00038 | 0.18 keV |
| 11.096 MeV 4$^+$ | 0.00439 | 0.00065 | 0.39 keV |

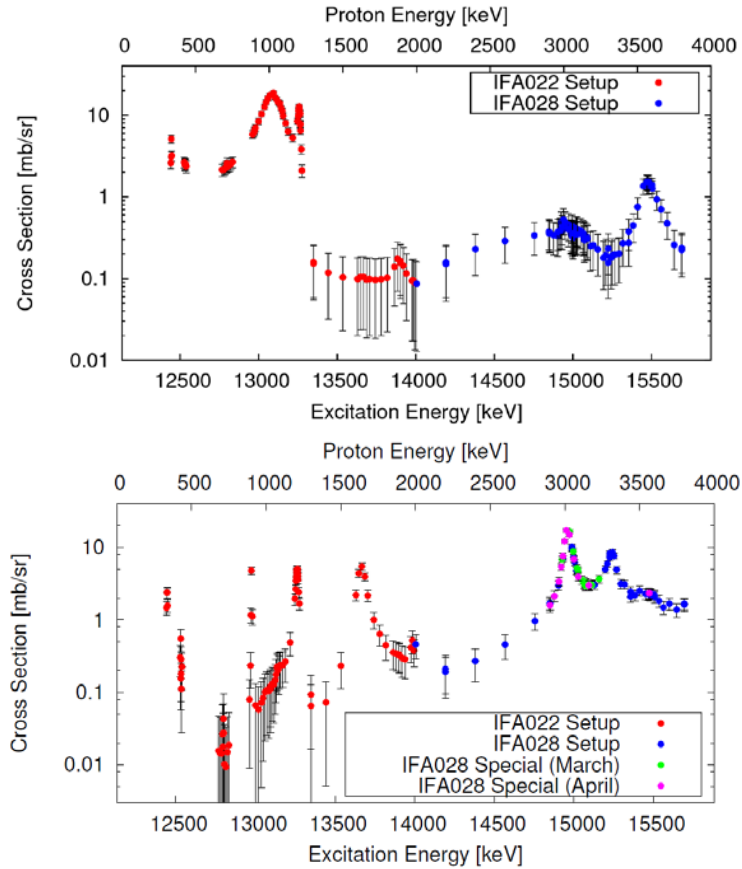

**FIGURE 3.** Normalized yield curves from the proton energy scan for the $^{15}$N$(p,\alpha_0)^{12}$C data.

In total, ten known excited states of $^{16}$O were identified from $E_x$ = 12 to 16 MeV. These are listed in Table 2, and they include the 13.091 MeV $1^-$ state which is in the alpha-0 spectrum at a proton energy of $E_p$ = 1.03 MeV. There is no evidence for the $E_x$ = 13.6 MeV $0_4^+$ state, possibly because it is obscured by the $E_x$ = 13.664 MeV $1^-$ state. The 15.046 MeV 2+/3- state from Ref. [8] cannot be seen in Fig. 3. This may be because the 14.92 MeV state, which is also known to have a very large alpha-1 branching ratio, may be interfering significantly in this energy region. There is clear evidence, however, at $E_x$ = 15.1 MeV of a broad state with a large alpha-0 strength which is presumably related to the $0_6^+$ state. These data will enable for an R-matrix description up to $E_x$ = 15.7 MeV in $^{16}$O, which may also improve the description at the astrophysically important lower energies.

**TABLE 2.** The known states in $^{16}$O which were observed in this measurement compared to their literature values. The uncertainty in the present measurement is related to small variations in beam energy.

| Literature value – $^{16}$O $E_x$ [MeV] | Reference no. | Current measurement – $^{16}$O $E_x$ [MeV] |
|---|---|---|
| 12.796 $0^-$ | [4] | 12.793(1) |
| 12.969 $2^-$ | [10] | 12.972(1) |
| 13.090 $1^-$ | [4] | 13.091(1) |
| 13.262 $3^-$ | [4] | 13.262(1) |
| 13.665 | [4] | 13.664(1) |
| 13.900 | [11] | 13.880(1) |
| 14.917 $2^+$ | [11] | 14.927(1) |
| 15.076 $0^+$ | [8] | 15.100(1) |
| 15.260 $2^+$ | [11] | 15.240(1) |
| 15.406 $3^-$ | [11] | 15.408(1) |

## CONCLUSION AND OUTLOOK

Through a scan in proton energy, and by extension in $^{16}$O excitation energy, the alpha-0 and alpha-1 breakup strengths of 10 known states of $^{16}$O could be determined. These include a candidate for the 4-alpha cluster state. The $^{15}$N($p,\gamma$)$^{16}$O reaction channels are obscured by interference from the $^{19}$F($p,\alpha$)$^{16}$O channels, but an upper limit could be estimated for their reduced widths as indicated in Table 1. A detailed R-matrix analysis of these states must be performed to extract their strengths and widths, and therefore their implications to stellar reaction rate models.

## ACKNOWLEDGMENTS

The authors are grateful to Jacques Chevallier and Folmer Lyckegaard for manufacturing the targets. We also acknowledge financial support from the European Research Council (ERC) starting grant LOBENA No. 307447. OSK acknowledges support from the Villum Foundation.